\documentclass[showpacs,preprintnumbers,amsmath,amssymb]{revtex4}
\usepackage{amsthm}
\usepackage{enumerate}
\usepackage{epsfig}

\newcommand*{\ket}[1]{|#1\rangle}
\newcommand*{\bra}[1]{\langle #1|}

\newcommand{\be}{\begin{equation}}
\newcommand{\ee}{\end{equation}}
\newcommand{\bea}{\begin{eqnarray}}
\newcommand{\eea}{\end{eqnarray}}
\newcommand{\bestar}{\begin{equation*}}
\newcommand{\eestar}{\end{equation*}}
\newcommand{\beastar}{\begin{eqnarray*}}
\newcommand{\eeastar}{\end{eqnarray*}}

\begin{document}

\title{The limits of counterfactual computation}

\author{Graeme \surname{Mitchison}}
\email[]{g.j.mitchison@damtp.cam.ac.uk} \affiliation{Centre for
Quantum Computation, DAMTP,
             University of Cambridge,
             Cambridge CB3 0WA, UK}

\author{Richard \surname{Jozsa}}
\email[]{richard@cs.bris.ac.uk} \affiliation{Department of Computer
Science, University of Bristol, Merchant Venturers Building, 
Bristol BS8 1UB, UK}

\begin{abstract}
We show that the protocol recently proposed by Hosten et
al. \cite{Hosten} does not allow all possible results of a computation
to be obtained counterfactually, as was claimed. It only gives a
counterfactual outcome for one of the computer outputs. However, we
confirm the observation \cite{Hosten} that the protocol gives some
protection against decoherence. In some situations, though, it may be
more effective simply to run the computer several times.

\end{abstract}
\date{\today}

\maketitle

\pagestyle{plain}

In their recent paper \cite{Hosten}, Hosten et al. proposed a novel
protocol, using the `chained Zeno' principle. This, it was argued,
would allow all possible results of a quantum computation to be
obtained, with probability close to 1, without running the
computer. There is, however, a limit to the information that can be
obtained from such a counterfactual computation. Consider a
computation with two outputs, 0 and 1, and let $p_i$ denote the
probability of inferring the result $i$ from a protocol without
running the computer, for $i=0 \mbox{ or } 1$. It was shown in
\cite{MitJozsa}, \footnote{Curiously, this is referred to as the
`random guessing limit' in \cite{Hosten}, though obtaining a
counterfactual result for one outcome, as in the special case of
interaction-free measurement, is very different from guessing.} that
$p_0+p_1 \le 1$. In defiance of this limit, the chained Zeno protocol
would allow $p_0+p_1$ to be arbitrarily close to 2 (in the case of a
computation with two possible outputs). The explanation for this
contradiction is that the protocol is not fully counterfactual, and it
is instructive to see why this is so.

\section*{The chained Zeno  protocol.}

We summarise the protocol, using the notation in the Supplementary
Material \cite{Hosten}. We consider a simplified computation where
there are only two outputs, 0 or 1. In other words, we consider only
two possible outputs rather than the four in their Grover search
example.

The protocol consists of a subroutine inside a routine. When the
subroutine runs, the computer can be activated if a switch qubit, the
`computer switch', is on (0=off, 1=on). We shall speak of an
`insertion of the computer' to mean a step of the subroutine where the
computer is activated if its switch qubit is on. There is a second
qubit, the `subroutine switch', that controls whether the subroutine
is entered from the routine. Finally, there is a third qubit that
receives the output of the computation. The entire state can be
written $\ket{\Psi}=\sum \alpha_{ijk} \ket{ijk}$, with
$$\ket{ijk}=
\ket{\mbox{subroutine switch=i}}\ket{\mbox{computer
switch=j}}\ket{\mbox{computer output=k}}.$$
Initially, $\ket{\Psi}=\ket{000}$.

Both the routine and the subroutine use the quantum Zeno principle
\cite{Misra}. At each step, the subroutine applies a rotation
$R=\left( \begin{array}{cc}
\cos \theta & -\sin \theta \\
\sin \theta & \cos \theta \\
\end{array} \right)$ (in the computational basis)
to the the computer switch, inserts the computer, then measures the
output qubit. There are $N$ steps in the subroutine, with $N
\theta=\pi/2$. The routine applies a rotation $R^\prime$ by an angle
$\theta^\prime$ to the subroutine switch, lets the subroutine run,
then measures the computer switch. There are $N^\prime$ steps in the
routine, with $N^\prime \theta^\prime=\pi/2$. 

After the first step of the routine, the rotation $R^\prime$, the
state is $\ket{\Psi}=\cos \theta^\prime \ket{000}+\sin \theta^\prime
\ket{100}$. Next, the subroutine is switched on in the second term
(but not in the first, where the subroutine switch is 0). After the
first step of the subroutine, the state is
\be \label{rotation2} \ket{\Psi}=\cos \theta^\prime \ket{000}+\sin
\theta^\prime \underbrace{\left(\cos \theta \ket{100}+\sin \theta \ket{110}\right)}_{subroutine}.  \ee
The brace here indicates the terms that change during the
subroutine. Next the computer is inserted, and runs in the second term
within the brace, where the computer switch is 1.

To see how the protocol works, we consider the two outputs separately:

\subsection*{Output 0}
When the computer runs, equation (\ref{rotation2}) is unchanged. This
happens at every step of the subroutine, which rotates to the final
state
\be \label{endofsub0} \ket{\Psi}=\cos \theta^\prime \ket{000}+\sin
\theta^\prime \ket{110}.  \ee
The next step in the routine is to measure the computer switch, and
this kills the second term with probability close to 1 if
$\theta^\prime$ is small. Thus we reach the state
$\ket{\Psi_0}=\ket{000}$, which is also the final state of the
routine. 

\subsection*{Output 1}

When the computer runs, the second term in the brace becomes $\sin
\theta \ket{111}$. With probability close to 1, this term is killed
when the output qubit is measured. At the end of the subroutine, after
$N$ steps, we have the (unnormalised) state $\cos \theta^\prime
\ket{000}+\sin \theta^\prime \cos^N \theta \ket{100}$, and, if $N$ is
large enough, and hence $\theta$ small enough, this is close to
$\cos \theta^\prime \ket{000}+\sin \theta^\prime
\ket{100}$. Measurement of the computer switch qubit does not change
this state, and the remaining steps of the routine rotate the state
fully, with probability close to 1, to the state $\ket{\Psi_1} =
\ket{100}$. This state is orthogonal to $\ket{\Psi_0}$; the two final
states can be distinguished, with probability close to 1, by measuring
the subroutine switch qubit at the end of the protocol.

\subsection*{Counterfactuality}

We now address the question of whether the computer runs during the
protocol. To make this precise, we keep track of whether the computer
has run by making a list of the possible sequences of measurement
outcomes that can occur, including in each list hypothetical
measurements of the computer switch qubit at the end of each insertion
of the computer. We call such a list a `history' \cite{MitJozsa}. Each
history $h$ is a list of measurement outcomes (real and hypothetical),
which correspond to a product of projectors $P_1 \cdots P_k$; we
associate to $h$ the un-normalised vector $v_h=P_1 \cdots P_k
\ket{000}$. As convenient notation, we write $0_i$/$1_i$ for the
outcome 0/1 of a measurement on the $i$-th qubit in the state, and
$f$/$n$ for an `off'/`on' outcome of the hypothetical measurement of
the computer switch. The history is then a list of outcomes and $f$ or
$n$ symbols, written left to right in the order of the protocol
steps. Tables \ref{histories0} and \ref{histories1} show the histories
and their associated vectors for one step of the routine, i.e. for the
first complete cycle of the subroutine followed by the computer switch
measurement. Here we take $N=2$.

\begin{table}
\caption{Histories after one subroutine cycle, with $N=2$. Output 0.
\label{histories0}}
\begin{center}
\begin{tabular}{|c|c|}
\hline
\hline
$h$ & $v_h$\\
\hline
$f0_3f0_30_2$ & $\cos \theta^\prime \ket{000}+\sin \theta^\prime  \ket{100}/2$\\
\hline
$f0_3n0_31_2$ & $\sin \theta^\prime  \ket{110}/2$\\
\hline
$n0_3f0_30_2$ & $-\sin \theta^\prime \ket{100}/2$\\
\hline
$n0_3n0_31_2$ & $\sin \theta^\prime  \ket{110}/2$\\
\hline
\hline
\end{tabular}
\end{center}

\caption{Histories for output 1
\label{histories1}}
\begin{center}
\begin{tabular}{|c|c|}
\hline
\hline
$h$ & $v_h$\\
\hline
$f0_3f0_30_2$ & $\cos \theta^\prime \ket{000}+\sin \theta^\prime \ket{100}/2$\\
\hline
$f0_3n1_31_2$ & $\sin \theta^\prime \ket{111}/2$\\
\hline
$n1_3f1_30_2$ & $-\sin \theta^\prime \ket{101}/2$\\
\hline
$n1_3n1_31_2$ & $\sin \theta^\prime \ket{111}/2$\\
\hline
\hline
\end{tabular}
\end{center}
\end{table}

If $m$ is a particular set of outcomes of the (actual) measurements
during the protocol, the histories containing $m$ are added coherently
to give the amplitude for $m$. Thus $|x|^2$ is the probability of the
outcomes $m$, where $x=\sum_{m \subset h} v_h$. We are now ready to define
a counterfactual computation:

A set $m$ of measurement outcomes is a {\em counterfactual outcome} if
(1) there is only one history associated to $m$ and that history
contains only $f$'s, and (2) there is only a single possible computer
output associated to $m$.

Looking at table \ref{histories1} for the output 1, we see that
condition (1) is satisfied for $m=0_30_30_2$ which occurs only in the
history $h=f0_3f0_30_2$. This is only a segment of a complete history,
covering only one step of the routine. But condition (1) also holds
for the complete set of measurement outcomes, and, as the final
subroutine switch measurement distinguishes $\Psi_0$ from $\Psi_1$,
condition (2) is satisfied. Thus $m$ is a counterfactual outcome for
the output 1.

The situation is different for output 0. Table \ref{histories0} shows
that there are two histories containing $m=0_30_30_2$, namely
$f0_3f0_30_2$ and $n0_3f0_30_2$. Since the latter contains an `n',
condition (1) is not satisfied, and there is non-vanishing amplitude
for the computer running when the set of measurement outcomes $m$ is
obtained. This is therefore {\em not} a counterfactual outcome for the
output 0. Another way of seeing this is to note that the first term in
the brace in equation (\ref{rotation2}), $\cos \theta \ket{100}$, is
removed during the subroutine by destructive interference from terms
where the computer runs; this occurs when we add $v_{f0_3f0_30_2}$ and
$v_{n0_3f0_30_2}$ in Table \ref{histories0}. Thus, for output 0, the
final state can only be reached if the computer has run during the
protocol.

Now let us consider the whole protocol, and let $m_i$ denote the
result of the final measurement of the protocol, with $m_0=0_1$,
$m_1=1_1$. Let $m$ denote the set of measurement results that must be
obtained up to that point for the protocol to succeed, i.e. a sequence
of $N$ $0_3$'s followed by an $0_2$ for each run of the
subroutine. Then, for instance, $P(mm_i|i)$ denotes the probability of
the protocol being successful and giving the outcomes appropriate to
output $i$ given that the actual computer output is $i$.

Let us define the {\em counterfactuality} of outcome $i$ by
$c_i=|v_{h^i}|^2$, where $h^i$ is the all-$f$ history for outcome
$mm_i$. Table \ref{countertable} gives the counterfactualities for a
choice of $N$, $N^\prime$ satisfying the condition $N \gg N^\prime$
for the protocol to run effectively \cite{Hosten}. The table also
shows the probabilities $P(mm_i|i)$. For a counterfactual outcome, we
expect $c_i=P(mm_i|i)$, since, by condition (1), only the all-$f$
history contributes to the outcome $mm_i$. For output 1, we do indeed
have $c_1=P(mm_1|1)$. However, for output 0, $c_0 \ll P(mm_0|0)$, and
we can infer that the outcome is far from counterfactual, since
histories in which the computer runs make the dominant contribution to
the measurement results.

\subsection*{Decoherence}

Suppose now that there is decoherence during the running of the
computer (but not during other parts of the protocol), and suppose
decoherence makes the computer act according to the admittedly
somewhat artificial rule:
\be \label{deco}
\ket{11x}\ket{e} \to (1-\epsilon)\ket{11x}\ket{e}+\sqrt{2\epsilon-\epsilon^2}\ket{11(1-x)}\ket{e^\prime},
\ee
where the computer output $x$ is 0 or 1, and $\ket{e}$ and
$\ket{e^\prime}$ are environmental states, assumed to be orthogonal,
and where $\ket{e^\prime}$ is a different state for each computer run.

Table \ref{decotable} shows the effect of decoherence with
$\epsilon=0.2$. The protocol used was not the chained Zeno protocol
described above, but a modification of it where the computer is
inserted twice in each step of the subroutine \footnote{The modified
protocol, described in the Methods section of \cite{Hosten}, has two
insertions of the computer instead of one at each step of the
subroutine, with a sign change of the $\ket{111}$ term in the state
between them. (At the second insertion, the inverse computation is
used, but this is the same as original computation in the case
considered here.) This means that for computer output 1, but not for
output 0, the rotation R is cancelled at each step. This protocol
can work efficiently even when $N \not\gg N^\prime$}; this works
efficiently over a wider range of values of $N$ and $N^\prime$. The
table shows $P(m|i)$, the probability of the protocol succeeding given
computer output $i$, and $P(m_i|mi)$, the probability of getting the
final measurement result appropriate to the computer output, given
that the protocol has been successful. Note that these probabilities
increase as the relevant counterfactuality increases. In fact, the
values of $N$ and $N^\prime$ chosen by Hosten et al. lie in the range
where the counterfactualities are approximately equal, as are the
relevant probabilities, and it is in this region that one gets the
best performance on both computer outputs.

There is therefore an intriguing hint of a connection between
counterfactuality and decoherence-resistance. We should bear in mind,
however, that the modified chained Zeno protocol inserts the computer
$2NN^\prime$ times, and time must be allowed on each insertion for the
computer to run or not run. We might therefore wish to compare the
protocol with the simpler procedure of just running the computer
$2NN^\prime$ times. Table \ref{inftable} shows the mutual information
gained by the chained Zeno protocol and by $2NN^\prime$ repeated runs
of the computer, and the latter wins except in the special
circumstance of a decoherence so large that the correct and incorrect
answers have roughly equal probabilities. Repeated runs then give
little information, whereas the interference of terms in the protocol
can distinguish between differing states of the environment.

The conclusion seems to be that the benefits of counterfactual
computation are limited, although special cases such as
interaction-free measurement \cite{Elitzur} may nevertheless offer
some genuine practical utility.

\begin{table}
\caption{`Counterfactuality'.
\label{countertable}}
\begin{center}
\begin{tabular}{|c|c|c|c|c|c|c|c|c|}
\hline
\hline
$N$ & $N^\prime$ & $c_0$ & $c_1$ & $P(mm_0|0)$ & $P(mm_1|1)$ \\
\hline
700 & 70 & 0.0015 & 0.884 & 0.965 & 0.884 \\
\hline
\end{tabular}
\end{center}

\caption{The modified algorithm of \cite{Hosten}, with $\epsilon=0.2$.
\label{decotable}}
\begin{center}
\begin{tabular}{|c|c|c|c||c|c|c|c|c|}
\hline
\hline
$N$ & $N^\prime$ & $c_0$ & $c_1$ & $P(m|0)$ & $P(m_0|m,0)$ & $P(m|1)$ & $P(m_1|m,1)$ \\
\hline
700 & 70 & 0.0015 & 0.884 & 0.609 & 0.040 & 0.973 & 0.9999 \\
\hline
40  & 70 & 0.188  & 0.175 & 0.625 & 0.975 & 0.630 & 0.969 \\
\hline
40  & 700 & 0.803  & 0.0042 & 0.948 & 0.9998 & 0.469 & 0.107 \\
\hline
\hline
\end{tabular}
\end{center}

\caption{Mutual information for two protocols.
\label{inftable}}
\begin{center}
\begin{tabular}{|c|c|c|c||c|c|c|c|c|}
\hline
\hline
$N$ & $N^\prime$ & $\epsilon$ & MI(Chained Zeno) & MI(Repeated runs) \\
\hline
10 & 10 & 0.2 & 0.46 & 0.9999 \\
\hline
2 & 2 & 0.2 & 0.324 & 0.360 \\
\hline
10 & 10 & $1-\sqrt{2}/2$ & 0.297 & 0\\
\hline
\hline
\end{tabular}
\end{center}
\end{table}

\begin{center}
\rule{8cm}{0.1mm}
\end{center}

\section*{Reply to the Response of Drs Hosten et al.}

In their response to us, Hosten et al. \cite{Hosten2} propose a new
definition of counterfactual computation. A key feature of this new
definition is that, if a set of histories have amplitudes that sum to
zero, then those histories can be discounted, and are not used in
assessing whether the computer runs. For instance, histories 1 and 2
in their Figure 2 have equal and opposite amplitudes and can
therefore be discounted.

This discounting rule seems very odd to us. In the classic two-slit
experiment, a dark fringe at a point P on the screen arises because
the amplitudes for a photon reaching P from the two slits sum to
zero. The usual view is that the photon goes through {\em both} slits
to reach P; their rule seems to imply that the photon goes through
{\em neither} slit on the way to P. Indeed, we can regard the two
slits as being analogous to the two arms of the interferometer in
their Figure 2, and it is only by arguing that the photon goes through
neither arm that they can claim that the computer C does not run.

We can highlight what is wrong with their new definition by adding a
fourth quantum register in their protocol that is initially set to
$\ket{0}$ and is incremented every time the computer runs. If the
computation is counterfactual, the computer never runs and therefore
the fourth register should always be zero. This is the case when the
computer output is 1, which we all agree is a counterfactual outcome.

Now consider what happens when the computer output is 0. In place of
their Table-I we get our Table \ref{counter}, and the amplitudes for
histories 1 and 2 no longer cancel. The protocol is therefore not
counterfactual according to their definition. But if it were
counterfactual without the fourth register, it should remain so with
this register added.

We conclude that the proposed new definition of counterfactual
computation is flawed, and we see no reason to depart from our own
definition.

\begin{table}
\caption{
\label{counter}}
\begin{center}
\begin{tabular}{|c|c|c|}
\hline
\hline
history & $h$ & $v_h$\\
\hline
1 & $n^{(1)}n^{(2)}0_3n^{(1)}f^{(2)}0_30_2$ & $-\sin\theta'\ket{1001}/2$\\
\hline
2 & $n^{(1)}f^{(2)}0_3n^{(1)}f^{(2)}0_30_2$ & $\sin\theta'\ket{1000}/2$\\
\hline
3 & $f^{(1)}f^{(2)}0_3f^{(1)}f^{(2)}0_30_2$ &  $\cos\theta'\ket{0000}$\\
\hline
\hline
\end{tabular}
\end{center}
\end{table}

\section*{Reply to further comments of Drs Hosten et al.}

In our Reply above, we introduced the idea of a fourth register that
counts the number of times the computer runs, and noted that the
chained Zeno protocol would not work with this register added. Hosten
et al.  are quite right to point out, in their Reply \cite{Hosten2},
that certain protocols we propose \cite{Jozsa99,MitJozsa} would also
not work with this fourth register added.  In fact, our definition of
a counterfactual computer output $r$ comprises two parts: the first
involves the protocol with computer $U_r$ (in which $U_r$ ``does not
run'') and the second involves the protocol with computer $U_{1-r}$
(in which the computer is allowed to run).  The fourth register above
spoils only the second part and not the first.

As a refinement of the fourth register idea, we can test the two
possible computer outputs individually. For instance, we can increment
the fourth register only when the output is 1, leaving it unchanged
for output 0. Then our protocol and the chained-Zeno protocol both
function correctly for the output 1, which is counterfactual.  As we
explained, that is what one expects, since the computer never runs and
the fourth register should therefore never change. But if one tests
the output 0 in this way on the chained-Zeno protocol, the
cancellation does not occur and the protocol fails. Yet it should not
fail if 0 were truly a counterfactual output.

We can state this more generally. Let us define a {\em tally register}
to be a quantum register that keeps a record in some way of whether
the computer runs when it has some specified output. It seems
reasonable to require, as a general property of counterfactual
computation, that adding a tally register for a counterfactual output
should not affect the protocol. 

This will always be true according to our definition of a
counterfactual computation. By contrast, if we use the definition of
Hosten et al., then whenever histories with cancelling amplitudes are
neglected there will be a tally register that makes the protocol
fail. This tally register is incremented in a way that depends on the
stage of the algorithm so that distinct histories have distinct final
entries in the register (this was why we talked about ``incrementing''
rather than flipping a bit). Clearly, if the computer runs in any of
the cancelled histories, cancellation no longer occurs and the
protocol fails. If the computer does not run in any of these
histories, it satisfies our definition and the cancellation was
unnecessary.

In the Appendix of \cite{Hosten2}, Hosten et al. argue that our
definition of counterfactuality is inconsistent. First they recall a
protocol originally defined in \cite{Jozsa99}. Here the first qubit is
the computer switch and the second receives the computer
output. Define $U$ by
$U\ket{0}_1\ket{0}_2=\ket{0+1}_1\ket{0}_2/\sqrt2$,
$U\ket{1}_1\ket{0}_2=\ket{1-0}_1\ket{0}_2/\sqrt2$ and
$U\ket{i}_1\ket{1}_2=\ket{i}_1\ket{1}_2$, for $i=0,1$. (Here we are
using the shorthand $\ket{0 \pm 1}$ for $\ket0 \pm \ket1$). Then the
protocol for the case where the computer output is 1 is as follows:
\be \label{computer1}
\ket{0}_1 \ket{0}_2 \ \  \underrightarrow{U}\ \  \ket{0+1}_1 \ket{0}_2/\sqrt 2 \ \  \underrightarrow{computer} \ \ (\ket{0}_1\ket{0}_2+\ket{1}_1\ket{1}_2)/\sqrt 2 \ \ \underrightarrow{U} \ \ \ket{0+1}_1\ket{0}_2/2+\ket{1}_1\ket{1}_2/\sqrt2.\ee
If a measurement of the two qubits yields $\ket{0}_1\ket{0}_2$, this
is a counterfactual outcome, according to our definition, because the
term does not arise when the computer output is 0, and because the
computer switch is never on (the first qubit is never set to
$\ket{1}_1$) in the terms that give rise to $\ket{0}_1\ket{0}_2$.

Hosten et al. then propose an internal structure for the computer,
replacing it by a Hadamard, followed by a $\pi$ sign-change for the
output 1, followed by a second Hadamard. Instead of using the computer
switch (the first qubit) to determine whether the computer is on, they
use the output qubit, arguing that this can only be set to 1 if the
computer runs. The computer step above is then expanded into:
\be \label{computer2}
\underrightarrow{H} \ \ \frac{\ket{0}\ket{0}}{\sqrt 2}+\frac{\ket{1}\ket{0+{\tilde 1}}}{2}  \ \ \underrightarrow{\pi} \ \ \frac{\ket{0}\ket{0}}{\sqrt 2}+\frac{\ket{1}\ket{0-{\tilde 1}}}{2}\ \ \underrightarrow{H} \ \ \frac{\ket{0}\ket{0}}{\sqrt2}+\frac{\ket{1}\ket{0+{1}}}{2\sqrt2}-\frac{\ket{1}\ket{\tilde 0- \tilde 1}}{2\sqrt2},
\ee
where a tilde marks terms that belong to a history in which the second
qubit has been set to 1 by the first Hadamard. The final operation of
$U$ now yields the state:
\[
\frac{\ket{0+1}\ket{0}}{2}+\frac{\ket{1-0}\ket{0}}{4}+\frac{\ket{1}\ket{1}}{2\sqrt2}-\frac{\ket{\tilde 1-\tilde 0}\ket{\tilde 0}}{4}+\frac{\ket{\tilde 1}\ket{\tilde 1}}{2\sqrt2},
\]
again with the tilde convention. We are not allowed to cancel the terms
in $\ket{1-0}\ket{0}$ since they have different histories in terms of
the setting of the second qubit. It is clear that, since
$\ket{0}\ket{0}$ occurs with a tilde, it is not a counterfactual
outcome according to the new definition of this term. This is the
supposed inconsistency: this outcome was counterfactual according to
our original definition, but with this alternative definition it is no
longer counterfactual. However, the rules have changed. We are using a
different qubit to assess whether the computer runs, and this new
criterion does not infallibly decide between ``on'' and ``off''. It is
not surprising that a different question gets a different answer. We
take this up again later.

\subsection*{The three-box paradox and weak measurement}

\begin{figure}
\centerline{\epsfig{file=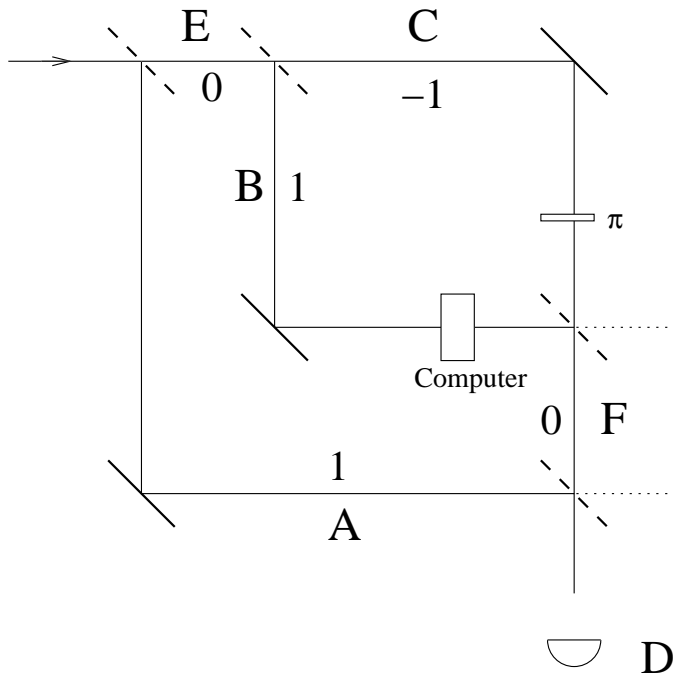,width=0.5\textwidth}}
\caption{The experiment considered by Vaidman \cite{Vaidman}, with the
labels he assigns to each path and the weak measurement results for
the photon being present on various paths in the case where the
computer output is 0. The beam-splitters at the top left and bottom
right transmit two-thirds of the beam and reflect one-third. Those in
the inner interferometer transmit and reflect equally.
\label{vaidman}}
\end{figure}

Since our last contribution to this discussion, there has been an
insightful contribution from L. Vaidman \cite{Vaidman}, which
demonstrates that a particular case of the chained-Zeno protocol is an
implementation of the ``three-box paradox''. In figure \ref{vaidman},
after the photon enters the inner interferometer, its state (labelled
by the paths shown in the figure) is
$(\ket{A}+\ket{B}+\ket{C})/\sqrt{3}$. Suppose the computer output is
0. Then, taking into account the phase-reversal $\pi$ and the effects
of the two subsequent beam-splitters, one sees that the photon will
end up in the detector D if the state is post-selected by
$(\bra{A}+\bra{B}-\bra{C})/\sqrt{3}$. Thus the claim by Hosten et al.
that the photon does not pass through the subroutine when the computer
output is 0 and detector D fires (see their discussion of their Figure
6) is equivalent to the claim that a photon, initially in state
$(\ket{A}+\ket{B}+\ket{C})/\sqrt{3}$ and post-selected to be in the
state $(\bra{A}+\bra{B}-\bra{C})/\sqrt{3}$, is not in ``box B''. The
argument in the case of the three-box paradox is that the photon must
be in box A because, if it were not, the state would collapse to
$(\ket{B}+\ket{C})/\sqrt{2}$, which has zero amplitude for being
post-selected. The paradox is that the same argument can be applied to
prove that photon must be in box B, which shows that the reasoning is
fallacious.

Vaidman observed that one can experimentally demonstrate that the
photon passes through the computer by carrying out a weak measurement
\cite{Aharonov98,Aharonov01}. A weak measurement of the projection on
B, post-conditioned on the photon being detected at D gives the value
1. For C one gets -1, and for F one gets 0. These weak measurement results
exactly parallel our argument that the photon must pass through the
computer in order for destructive interference to occur at the output
of the inner interferometer.

%This
%measurement leaves the photon's state largely undisturbed and carries
%only a small amount of information about this state. Nevertheless, an
%accurate result can be obtained by averaging over many repeats.

%The value $w_M$ of a weak measurement of $M$ is given by
%
%\be \label{weakformula}
%w_M=\frac{\bra{\phi_f}VMU\ket{\phi_i}}{\bra{\phi_f}VU\ket{\phi_i}},
%\ee
%
%where $\ket{\phi_i}$ and $\ket{\phi_f}$ are the initial and final
%states of the system, and $U$ defines the evolution of the state to
%the point where the measurement of $M$ occurs and $V$ defines the
%further evolution to the final state. Inserting $M=P_B$, i.e. the
%projection onto $B$, and
%$U\ket{\phi_i}=(\ket{A}+\ket{B}+\ket{C})/\sqrt{3}$,
%$\bra{\phi_f}V=(\bra{A}+\bra{B}-\bra{C})/\sqrt{3}$, we find $w_B=1$,
%confirming that the photon is present. Weak measurement for the
%

%When the computer is output is 1, we can defer the measurement of the
%output qubit by entangling the photon's state with an additional
%qubit, so the interaction with the computer takes the form
%
%\[
%(\ket{A0}+\ket{B0}+\ket{C0})/\sqrt{3} \to
%\ket{A0}+\ket{B1}+\ket{C0})/\sqrt{3}.
%\]
%
%We then post-select $(\bra{A0}+\bra{B0}-\bra{C0})/\sqrt{3}$. Weak
%measurement of the projection on $B$ then gives $w_B=0$, again showing
%that weak measurements agree with our definition of counterfactual
%computation.

\subsection*{The critique of weak measurement by Hosten et al.}

Hosten and Kwiat \cite{Hosten-weak} have responded to Vaidman's
analysis by questioning the meaning of the weak measurement results
shown in Figure \ref{vaidman}. First they point out that a weak
measurement could be obtained by inserting a tilted parallel glass
slab into the path to be measured; this slightly shifts the transverse
spatial distribution of the photon and this shift can be read out on
path D. If this is done on path B, ``there is no longer perfect
interference on path F'', and the resulting ``leaking amplitude'' on
path F causes the shift on path D.

We agree with this description. However, they then say that the
photons seen in this weak measurement ``come from the computer {\em
because} of the weak measurement'', and the photons are only ``weakly
present on path B''. Their claim, therefore, is that weak measurement
causes photons to pass through the computer, which they would not have
done without this measurement taking place. If this is so, what is the
block of glass deflecting? A photon that wasn't there until the block
was inserted?

\subsection*{Inconsistency re-examined}

We now return to the question of the supposed inconsistency in our
definition. The protocol given by the steps (\ref{computer1}) gives a
counterfactual outcome for the output 1 if we follow our definition
and use the computer switch to decide if the computer is on. This is
confirmed by weak measurement of the computer switch, which gives the
value 0 for the projection onto $\ket{1}_1$. However, if we follow
Hosten et al. in representing the computer by the sequence of steps
(\ref{computer2}), and if we deem the computer to be ``on'' if the
output qubit is set to 1, then the outcome is no longer
counterfactual. If this is inconsistent, then so are the laws of
nature, for if we do a weak measurement of the projection onto
$\ket{1}_2$ we get the non-zero answer $1/\sqrt{2}$.

%For measurement of $\ket{1)_2$ by P
%w=<00| VP | 00/sqrt2 + 1(0+1)/2>/<>
%where V is the inverse of
%00 -> (0+1)0/sqrt2
%01 -> 01
%10 -> (1-0)0/2 + 11/sqrt2
%11 -> -(1-0)0/2 + 11/sqrt2
%
%For measurement of $\ket{1)_1$ by P
%w=<00| U`P | (00+11)/sqrt2>/<>
%where U` is the inverse of U given by (see above):
%00 -> (0+1)0/sqrt2
%01 -> 01
%10 -> (1-0)0/sqrt2
%11 -> 11
%

The situation is, in fact, closely analogous to that described by
Vaidman, who notes that weak measurement of the projections onto both
E and F give zero (Figure \ref{vaidman}), whereas the weak measurement
result for C is 1. As he says: ``The photon did not enter the
interferometer, the photon never left the interferometer, but it was
there!''  \cite{Vaidman}. The interferometer here is the counterpart
of Hosten et al.'s internal structure for the computer, interpreting
the Hadamards in (\ref{computer2}) as the operation of
beam-splitters. Their example does not impugn our definition; instead,
it affirms that the answer one gets depends on the question one asks.

By contrast, the definition of counterfactuality that Hosten et
al. offer does seem to suffer from internal inconsistency, as pointed
out by Dr Finkelstein \cite{Hosten}, because there may be an ambiguity
about which histories should be cancelled. They consider an example
that corresponds to our Figure \ref{vaidman}. There are three
histories, following paths B, C and A, that have amplitudes $a$, $-a$
and $a$, respectively (in their Figure 8 they call these histories 1,
2 and 3, respectively). It seems that we could at whim cancel either
the first and second of these histories, or the second and third. They
argue that the cancellation of B and C happens first in passing
through the apparatus, and therefore this is the ``correct''
cancellation. Thus, they say, the detection of the photon at D must be
entirely due to the history through A.

In terms of Vaidman's three-box analogy, this says that the photon
must be in box A. However, as Vaidman observes, one can equally well
argue that the photon must be in box B. This makes it clear that their
choice is essentially arbitrary. Weak measurements bear this out,
because these give $w_B=1$, $w_C=-1$ and $w_A=1$, showing that the
ambiguity about cancellation is mirrored in physical data.

%Finally note that weak measurement of path B by the mechanism they
%suggest causes a slight perturbation so the the amplitudes for the
%histories along B and C no longer cancel. Accordingly, their rule
%should force them to cancel and discount the histories through A and
%C, as these are now the only histories with equal and opposite
%amplitudes. Before the perturbation, histories B and C were cancelled,
%but now, with the perturbation, a photon detected at D suddenly comes
%exclusively from path B via the computer. An epsilon difference and
%the whole story has changed!

\subsection*{Conclusion}

Our arguments and Vaidman's \cite{Vaidman} lead to the same
conclusion: that the chained Zeno algorithm is not counterfactual on
all its outputs. Perhaps the most convincing evidence comes from weak
measurement. This predicts that, if we select experiments where the
photon is detected at D, then a pointer weakly coupled to the path
containing the computer will show a deflection (when averaged over
many repeats of the experiment). We believe that the argument in
\cite{Hosten-weak}, that this deflection is {\em caused} by the weak
measurement, is incorrect, and indeed it would be remarkable if a
perturbation on a path where no photon was present caused a photon to
appear.

It seems to us that the fundamental flaw in their concept of
counterfactuality is contained in a sentence in \cite{Hosten-weak},
where they correctly point out that there is zero amplitude for a
photon to travel down path F, but then infer that ``a photon does {\em
not} pass through path B, or the computer, before arriving at path
D''. In experiments where a photon is detected at D, there is a
non-zero amplitude for it passing along B and a non-zero amplitude for
it passing along A, and the fact that there is destructive
interference where paths B and C meet does not mean that the amplitude
for path B somehow cannot get to the detector and is thereby rendered
non-existent. Instead, we have a superposition where the amplitudes
for paths A and B both contribute to the detector at D firing. This is
beautifully demonstrated by Vaidman's three-box analogy.

\subsection*{Acknowledgements}

We thank Prof. D.J.C. MacKay and Prof. S. Popescu for helpful discussions.

\end{document}